# A comparative study of CO adsorption on flat, stepped and kinked Au surfaces using density functional theory

Faisal Mehmood[1], Abdelkader Kara[2,*], Talat S. Rahman[2], and Claude R. Henry[3,4]

[1]*Material Science Division, Argonne National Laboratory, 9700 South Cass Avenue, Argonne, IL 60439*
[2]*Department of Physics, University of Central Florida, Orlando, FL32816, USA*
[3] *Centre Interdisciplinaire des Nanosciences de Marseille, UPR CNRS 3118, 13288 Marseille Cedex 09, France*
[4] *associated to Aix-Marseille Universite, France*

**ABSTRACT**

Our *ab initio* calculations of CO adsorption energies on low miller index (111), (100), stepped (211), and kinked (532) gold surfaces show a strong dependence on local coordination with a reduction in Au atom coordination leading to higher binding energies. We find trends in adsorption energies to be similar to those reported in experiments and calculations for other metal surfaces. The (532) surface provides insights into these trends because of the availability of a large number of kink sites which naturally have the lowest coordination (6). We also find that, for all surfaces, an increase in CO coverage triggers a decrease in the adsorption energy. Changes in the work-function upon CO adsorption, as well as the frequencies of the CO vibrational modes are calculated, and their coverage dependence is reported.

[*]Corresponding author: kkara@physics.ucf.edu

## I. INTRODUCTION

Investigation of the adsorption of molecules on well-defined transition metal surfaces has been of great academic as well as technological interest for the past few decades [1-3]. In this regard, the focus has been mostly on CO adsorption due to its importance in many industrial processes and to its relative simplicity, contributing to the understanding of catalytic reactions [4-6]. The CO adsorption on a variety of transition metal surfaces has been reported vastly in the literature both experimentally and theoretically [7, 8]. The importance of identifications of 'active sites' based on adsorption, desorption and sticking coefficients is emphasized in many papers [1-3], as real catalysts are known to consist of small metal clusters of various micro-facets of different orientations containing defects like steps and kinks [9]. Numerous studies have attempted to explain unexpected catalytic activity of CO oxidation on nano-sized gold clusters on supported metal oxides. These supported nano-clusters exhibit very similar behavior to single crystal surfaces having steps and kinks as there is a large number of step and kink-like sites present on these nano-clusters. It has been reported that oxygen molecules dissociate more easily on step and kink sites and CO binds more strongly to such sites. Hence, systematic information on the adsorption characteristics of CO on Au surfaces with varying number of step and kink sites as a building block in advancing knowledge of the reactivity of metal particles [10]. Although recent thermal desorption spectroscopy (TDS) studies of CO on various surfaces of copper provided some insight of the dependence of binding energies on coordination [10], analysis based on coordination alone does not provide a detailed picture, as has been reported earlier [11]. We believe that a more in-depth understanding can be developed by systematic and detailed theoretical calculations using first principles methods as in density functional theory. Even though, there are number of calculations already reported for various coverage of CO on low miller index (111), (100), and (110) surfaces of transition metals, unfortunately we were unable to find studies dealing with exact same experimental coverage [12, 13]. More recently TPD experiments of CO adsorption on low-coordinated sites on Au(211) and Au(332) [14, 15] provide insight on the role of step and kink sites showing increased binding energies. Earlier calculations show that CO binding energy increases on Au(211) compared to Au(111) [16, 17]. Yim et al also show an increase in CO binding energy on Au(332) with local coordination by artificially introducing defects on Au(332) [15]. To our knowledge, there are no published data or calculations for kinked gold surfaces, though this kind of data has been published for other transition metal surfaces [10].

The chemisorption of CO on transition metal surfaces has been of great discussion due to its importance in many catalytic processes but recently, renewed attention is drawn due to the fact that theoretical calculations show adsorption site preferences different from those experimentally observed [13]. The goal of the present work is to calculate adsorption geometries, structural properties and electronic structure of CO adsorbed on experimentally observed sites on various flat, stepped and kinked surfaces.

In this paper, we present a detailed density functional study of CO adsorption on low Miller index ((111) and (100)), stepped (211) and kinked (532) surfaces of gold. After giving the computational details of our work in the next section, we shall first give a detailed analysis of the calculated adsorption energies based on coordination with comparison to other calculations and experiments. We then explain the implication of increasing CO coverage on these surfaces. The remaining of the paper is devoted to the characterization of these surfaces on basis of electronic structure (work-function, and vibrational frequencies) and the change in these quantities due to the adsorption of CO as compared to the clean surface followed by our conclusions.

## II. COMPUTATIONAL DETAILS

*Ab initio* calculations performed in this study are based on density functional theory (DFT) [18, 19]. A comprehensive study of energetics and electronic structure was made by solving Kohn-Sham equations in the plane-wave basis set using Vienna *ab initio* simulation package (VASP) [20-22]. The electron-ion interaction for C, O, and Au is described by ultra-soft pseudo-potentials supplied with VASP. A 450 eV plane-wave energy cut-off was used for all calculations and is found to be sufficient for these systems as reported by other studies [13, 23]. In all calculations, the generalized gradient correction of Perdew and Wang [24] (PW91) was used which is found to give more accurate results than those based on the local density approximation (LDA) [12, 25, 26]. The bulk lattice constant was found to be 4.19 Å using a k-point mesh of 10×10×10.

The slab supercell approach with periodic boundaries is employed to model the surface with the Brillouin zone sampling based on the technique devised by Monkhorst and Pack [27]. Au(100) and Au(111) were modeled by a 4 layer (16 atoms) tetragonal and hexagonal supercells, respectively. These 4 layers are separated with 11 Å of vacuum. For Au(100), calculations were performed for a c(2×2) overlayer corresponding to 0.5 ML coverage of CO, followed by calculations with full coverage. For Au(111), a p(2×2) structure was used along with other coverages ranging from 0.33 ML (1 CO per unit cell) to 1 ML(3 CO molecules per unit cell). CO molecules were adsorbed on several sites to find the preferred adsorption site in such a way that CO molecule sits perpendicular to the surface with carbon atom close to the surface as reported in number of experiments [7, 22, 23]. A Monkhorst-Pack k-point mesh of 4×4×1 was used for (100) and 5×5×1 was used for (111).

The stepped surface, Au(211), was modeled by an orthorhombic supercell of 17 layers separated with approximately 12 Å vacuum. Au(211) is a surface with a monatomic (100) step and a (111) 3-atom wide terrace (see Figure 1c). For this surface, a (2×1) unit cell was modeled with 34 Au atoms (2 atoms long steps) with CO adsorbed on step edge and on bridge (between two step edge atoms) sites. This supercell corresponds to 0.17 ML coverage on this surface. Calculations for a higher coverage of 0.33 ML were also performed and were achieved by incorporating an additional CO molecule to the same supercell to determine the coverage dependence of adsorption energies. A Monkhorst-Pack k-point mesh of 5×4×1 was used for Au(211).

Finally, the kinked surface, Au(532), (see Figure 1d) was modeled by a simple monoclinic supercell of five layers with each layer containing 8 non-equivalent atoms. These five layers were separated by 12 Å of vacuum. A Monkhorst-Pack k-point mesh of 3×4×1 was used and CO was adsorbed on the kink site and three bridge sites between atoms 1 – 4. The top of the kink site was found to be the preferred site which is also an experimentally observed preferred site for other transitions metals [10]. Two different coverages for the (532) surface were modeled by adsorbing only one CO molecule on the kink site and for the lower coverage an additional CO molecule on a site very next to the kinked site, for the higher coverage.

For all surfaces, each atom was allowed to move in all three directions and the structures were relaxed until forces on each atom were converged to better than 0.01 eV/Å. The adsorption energies were calculated by subtracting the energies of a CO-molecule in the gas phase and a clean-Au surface from the total energy of CO/Au system

$$E_{ad} = E_{CO/Au} - E_{CO} - E_{clean} \qquad (1)$$

Work-functions were calculated by taking the difference of average vacuum potential and the Fermi energy for each surface. Finite difference method was used to obtain vibrational properties of CO molecule in gas phase and on the surfaces. CO stretching and CO-surface frequencies were calculated in the direction perpendicular to the surface.

## II. RESULTS AND DISCUSSION

To understand trends in adsorption energies as a function of variation in the coordination of surface atoms, we have performed calculations on the low miller index Au surfaces ((111) and (100)), which have already been explored both experimentally and theoretically [10, 13, 28-30]. We have then extended these calculations to the stepped surfaces Au(211) and a kinked Au(532) surface for which, to our knowledge and unlike the copper case [31], no experimental data are so far available.

### III.1. Adsorption sites and bond length

In Table 1, we have compared our calculated values of CO bond length on various surfaces and the surface-carbon distance. As in all cases CO is adsorbed on top site of each surface and we note a very much expected similarity of CO bond length and surface-carbon distance. These distances are very close to experimentally observed values and to the ones calculated by others [10, 13, 28-30].

### III.2. Adsorption energies and their coverage dependence

An atom on a (111) surface has the highest number of nearest neighbors (coordination 9) and in turn lowest adsorption energy of all studied surfaces. For this particular surface, we have used a *p*(2×2) cell which corresponds to 0.33 ML CO coverage on Au(111). The calculated adsorption energies are listed in Table 2 along with corresponding number of first and second nearest neighbors and the experimentally observed adsorption energy. We studied several possible adsorption sites (top, bridge, hcp-hollow and fcc-hollow) and find only an energy difference less than 0.01 eV between them. Our calculated value of 0.28 eV is lower than the experimentally observed value of 0.40 eV, but it is comparable with other theoretical calculation for other noble metal surfaces (the adsorption energies obtained with DFT are often underestimated) while the reverse is true for most transition metal surfaces[13]. Our calculated value matches well with those obtained in other calculations performed for 0.25 ML and 0.33 ML coverage with reported adsorption energy of 0.32 eV and 0.30 eV, respectively [13, 16]. We have calculated adsorption energies with various CO coverages and for an accurate comparison, the same supercell was used to calculate different coverage for any given surface. To calculate higher coverage, extra CO molecules were adsorbed on the same supercell. In general, we found a strong dependence of the adsorption energies on coverage which was quite substantial especially for Au(111) for which the adsorption energy dropped from 0.28 eV to 0.1 eV (corresponding to 0.33 and 1ML, respectively). This also explains the small difference from other calculations which were performed for a smaller coverage and in turn larger adsorption energy.

The next surface in this hierarchy is Au(100) on which atoms have 8 nearest neighbors. Unlike the (111) surface, there are only a few calculations and experiments have been performed on this surface [30, 32, 33]. We have studied a c(2×2) coverage which corresponds to a 0.5 ML CO coverage on Au(100). Of the three possible adsorption sites for this surface we found the bridge site to be slightly energetically preferable over the top site with an energy difference of 0.04 eV. We obtained an adsorption energy of 0.38 eV for Au(100) which is larger than that on the highly coordinated surface (111). This value is again lower compared to the experimental value of 0.601 eV obtained using electron energy loss spectroscopy (EELS) [34]. Although, there are no other calculations available for this coverage a much smaller adsorption energy was reported for 1 ML coverage, in agreement with our high coverage calculations which show the strong drop in adsorption energy with increasing coverage.

The step atoms on Au(211) have coordination 7, and hence are next in order in this series. We have used a (2×1) cell for (211) corresponding to about 0.17 ML CO coverage. The CO molecule is adsorbed on the top and bridge sites of the step edge of (211), as shown in Figure 1c with adsorption energy of 0.54 eV. There are no experimental results available for this particular surface but TDP study on another stepped surface, Au(332), showed 0.57 eV for the adsorption energy which is reasonably close to our calculated value [14].

CO adsorption on Au(532) surface is studied on the kinked atom. The adsorption energy for this case was found to be 0.68 eV in our calculations which is the same as for the previous surface with coordination 7. This kind of similarities in adsorption energies of (211), and (532) has been seen experimentally for Cu surfaces where the authors have reported almost the same energies for these surfaces in their thermal deposition spectroscopy (TDS) data. Figure 2 shows the trends in adsorption energies versus CO coverage on the presently studied gold surfaces. Note that for each case CO binding energies decreases with increasing CO coverage.

### III.3. Vibrational properties and work-function

The vibrational frequency of a CO molecule in the gas phase was calculated by fully relaxing a single molecule in a large super cell of the size of approximately 6 ×6 × 22 Å.

The frequency of the stretching mode was calculated to be 2132 cm$^{-1}$. This can be compared with the very accurate experimentally measured vibrational frequency of an isolated CO molecule of 2079 cm$^{-1}$ using electron energy loss (EELS) and 2080 cm$^{-1}$ using infrared (IR) spectroscopy [35, 36]. For CO-covered Au surfaces, we have calculated two modes: the surface – molecule ($\nu_{Au-CO}$) and the molecule stretching frequency ($\nu_{CO}$). We find a small drop in the vibrational frequency compared to that for the free-CO as a result of the bonding of CO with Au surface atoms. The frequencies for all surfaces are summarized in Table 3. The small decrease in CO-stretch can be attributed to the strength of CO bonding to the gold surface. As shown in Figure 3, the CO stretching frequency decreases as CO binding energy increases since the stronger CO bond does not allow the molecule to vibrate faster although the differences are not substantially large.

We find a small increase in the surface-CO frequency ($\nu_{Au-CO}$) which we correlate to the decrease in coordination of the substrate surface. The stronger effect can be seen on the bridge site where the CO stretching frequency drops drastically due to an increase in coordination. In Table 3, we have also reported CO vibrational frequencies for higher CO coverage on gold surfaces. We found a significant decrease in vibrational frequencies with increase in CO coverage that can be understood on the basis that as CO coverage increases the CO-CO interaction comes into play.

We have extracted the work-function for all the surfaces mentioned above with and without CO molecules and our results are summarized in Table 4 along with the available experimental values for clean surfaces. Our calculated work-functions are smaller than the experimental values [37-39]. Our calculated values show a systematic decrease in the work-function for clean surfaces with decrease in coordination of substrate atoms. A similar behavior has been seen in experimental studies for (111), (100) and (110) surfaces [37-39] and theoretical studies on other metal surfaces [37-39]. Our calculations also show an increasing trend on work-function when CO is adsorbed on the surface with the strongest effect being for Au(100) surface with a work-function change of 0.57 eV while for the other surfaces this change is below 0.17 eV.

### III.4. Local electronic structure

Previous investigations pointed out to substantial –coordination dependent- alteration of the substrate electronic structure upon CO adsorption [31, 44]. Since our present study involves adsorption of CO in a variety of structural environments on Au surfaces, we present the changes in the local electronic densities of states (LDOS) for atoms on a step and a kinked surface on which CO is adsorbed on the top and the bridge sites. In Figs 4a and b we show the LDOS for a Au atom at the step edge (4a) and at the kinked site (4b) along with that of the oxygen and carbon atoms. For the case of CO/Au(211), we note a strong shift (as much as 2 eV) in the d-band towards stronger binding, along with a substantial narrowing of the d-band when the CO is adsorbed at atop site reflecting the a strong coupling between the electronic structures of CO on one hand and the gold step atom on the other. The shift as well as the narrowing is also noticeable for the case of CO adsorption at the bridge site but are not as dramatic as for the case of atop adsorption. Another feature worth mentioning is the *sp-d* hybridization at about 8.5 eV binding energy which is present in the atop adsorption case and absent in the bridge case. However, a weaker hybridization is noticed for both cases at about 7.5 eV below the Fermi level. Similar features are observed for the case of CO/Au(532) where the kink atom's experience the same changes in the electronic structure as the step atom for the same adsorption pointing to the fact that alterations to the electronic structure due to reduction in the coordination are not linear and saturate for low coordination (here a kink atom has a coordination 6 while a step atom has a coordination 7).

## IV. CONCLUSIONS

In this paper, we have presented results of theoretical investigations of CO adsorption on two low miller index surfaces, a stepped, and a kinked gold surface. The CO molecules were adsorbed on several sites on these surfaces to calculate preferred adsorption sites and to explain trends in adsorption energies resulting from differences in local geometrical and electronic structure. For all cases, we found that the CO adsorption energy depends strongly on coverage and drops with increase in coverage. WE also find the adsorption energy to increase with decrease in local coordination of surface atoms for low miller index surfaces and to saturate for the low coordinated (stepped and kinked) surfaces, which are the most favorable surfaces for CO adsorption. A very small drop in the vibrational frequency of the free CO molecule was noted when it was adsorbed on the metal surface but differences within the set of surfaces was found to be negligible for on-top adsorption, but significant for other sites. Work-function of adsorbate covered surfaces was found to decrease in all cases as compared to the clean surfaces.

**AKNOWLEGMENTS**

AK thanks CINaM for hospitability and support. We also acknowledge financial support from NSF Grant No. CHE-0741423.

**Table 1** Calculated structural properties of CO adsorbed on experimentally reported preferred adsorption sites on various Au surfaces. Here $d_{C-O}$ is the CO bond length and $d_{C-Au}$ is the CO-molecule distance from Au-surfaces. All distances are in Å.

| Surface | Au(111) | Au(100) | Au(211) | Au(532) |
|---|---|---|---|---|
| $d_{C-O}$ | 1.14 | 1.16 | 1.15 | 1.15 |
|  | 1.15, 2.18** | 1.17* | 1.17* |  |
| $d_{C-Au}$ | 2.22 | 2.03 | 1.99 | 2.00 |
|  |  | 1.50* | 1.46* |  |

*Bridge site ** Ref [16, 17]

**Table 2** Calculated adsorption energies of CO on various Au surfaces for the on top site (unless otherwise stated).

| Surface |  | Au(111) | Au(100) | Au(211) | Au(532) |
|---|---|---|---|---|---|
|  | $N_{NN}$ | 9 | 8 | 7 | 6 |
|  | $N_{2NN}$ | 3 | 5 | 3 | 3 |
|  | 1CO/Cell | 0.28 | 0.38 (0.46*) | 0.54 (0.65*) | 0.68 |
| $E_{ad}$ (eV) | 2CO/Cell | 0.16 | 0.17 | 0.46 | 0.62 |
|  | 3CO/Cell | 0.1 | - | - | - |
| $E_{ad}$ (eV)-exp |  | 0.4** | 0.6** | 0.52*** |  |

*Bridge site
** Ref [40]
***Ref [41]

**Table 3** Calculated CO frequencies: $\nu_{C-O}$ represents the molecular stretching mode and $\nu_{S-CO}$ is the molecule surface stretch mode. $CO_{mol}$ = 2132 cm$^{-1}$

| Surface | Au(111) | Au(100) | Au(211) | Au(532) |
|---|---|---|---|---|
| $\nu_{C-O}$ (cm$^{-1}$) | 2046 | 2040 | 2034 | 2035 |
|  | 2004* | 2019* | 2012* | 2034* |
|  |  | 1902** |  |  |
| $\nu_{S-CO}$ (cm$^{-1}$) | 255 | 278 | 288 | 314 |

*Higher CO coverage
**Bridge site

**Table 1** Calculated work-function for clean and CO adsorbed Au surfaces along with available experimental values.

|  | Au(111) | Au(100) | Au(211) | Au(532) |
|---|---|---|---|---|
| Clean | 5.11 | 5.14 | 5.09 | 5.03 |
| CO/Au | 4.55 | 4.51 | 4.71 | 4.67 |
| Clean$^{EXP}$ | 5.26-5.55 | 5.22 |  |  |

*Ref [38, 42, 43]

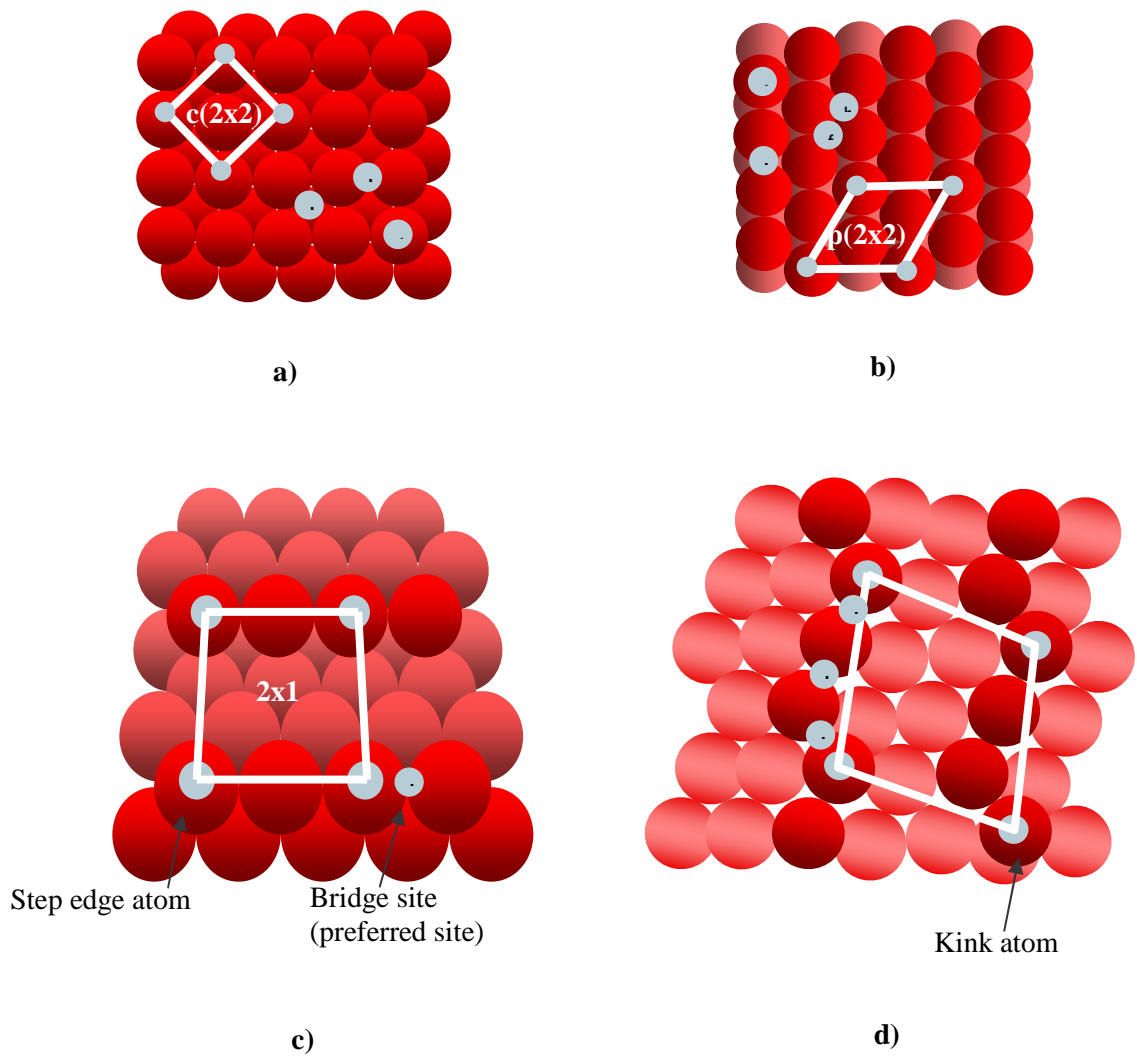

Figure 1 Top view of: a) (100), b) (111), c) (211), and d) (532) surfaces. Adsorption on different sites is shown. Dark colors represent top layer and CO molecules are represented by small circles. Note that in unit cell of (532) eight atoms belong to eight different layers.

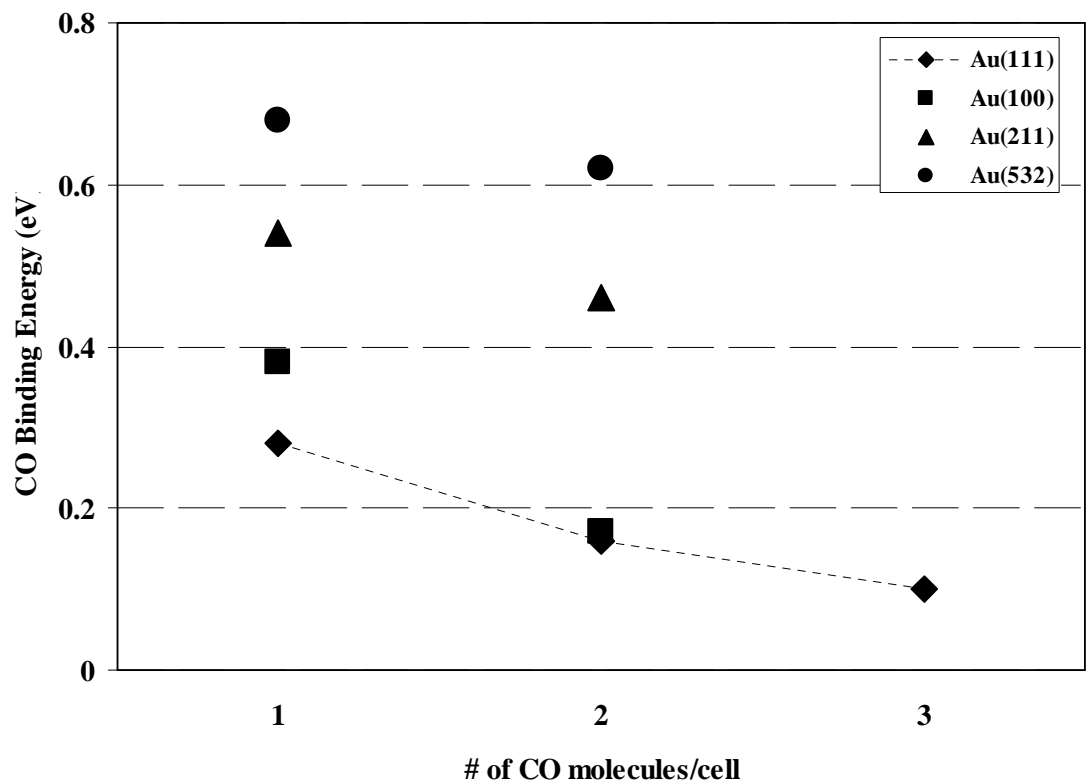

**Figure 2 CO binding energy as a function of CO coverage.**

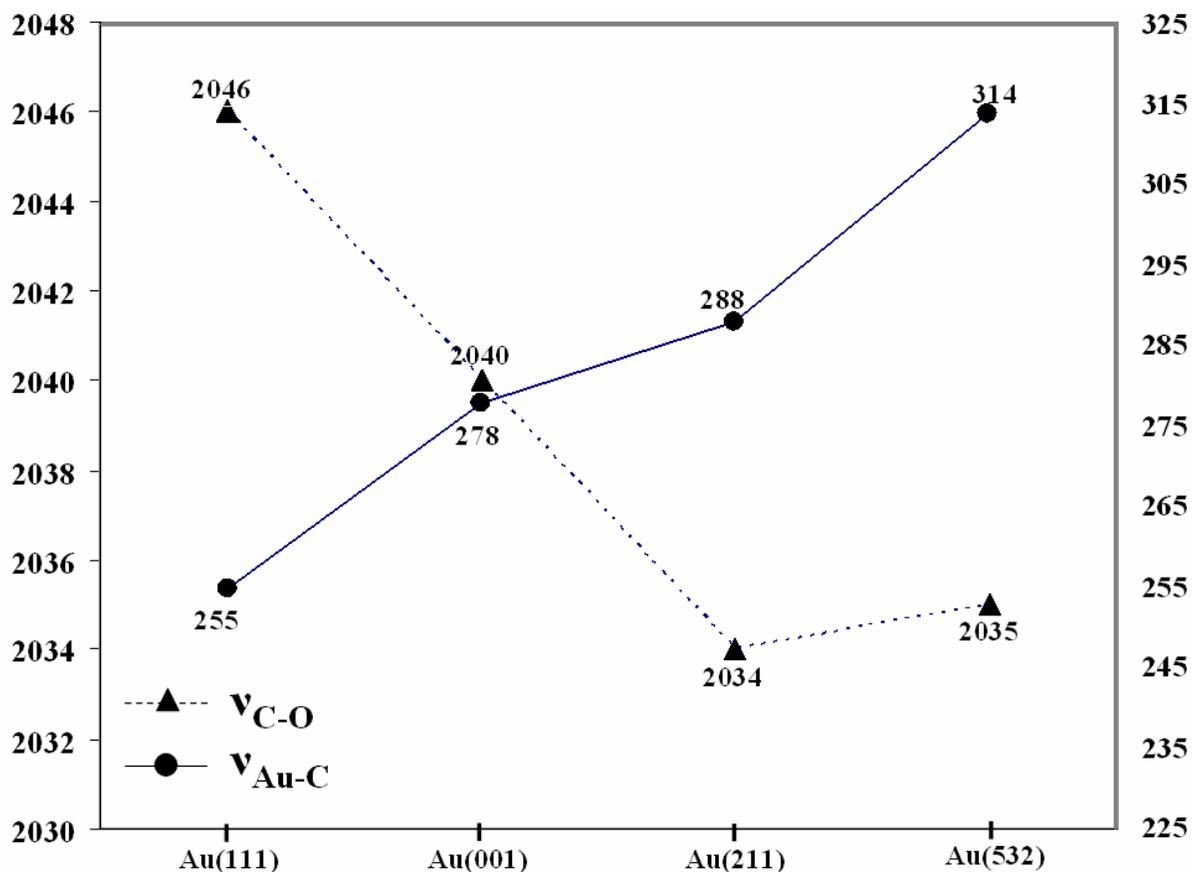

**Figure 3 C-O and Au-CO vibrational frequencies.**

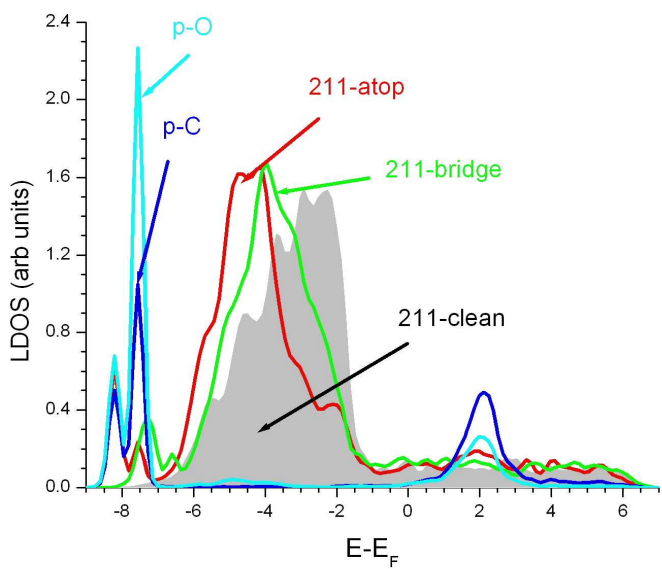

(a)

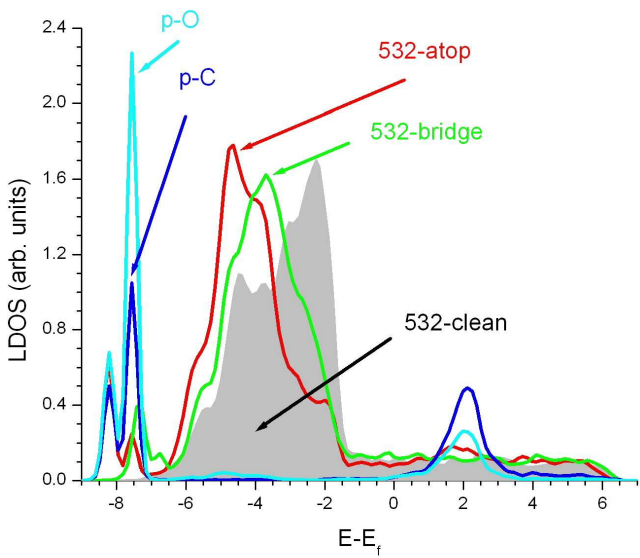

(b)

**Figure 4:** Local densities of states for Au, O and C atoms for the cases of a) CO on Au(211) and b) CO on Au(532).